\documentclass[preprintnumbers,aps,prd,showpacs,floatfix,superscriptaddress,nofootinbib,letterpaper,twocolumn]{revtex4-2}
\usepackage[utf8]{inputenc}
\usepackage{graphicx} 
\usepackage{xcolor}
\usepackage{bm}
\usepackage{amsmath, amsthm, amssymb}

\usepackage{hyperref}
\usepackage[top = 2cm, bottom = 2cm, right = 2cm, left =2cm]{geometry}
\begin{document}

\title{The scalar--Maxwell--\texorpdfstring{$\Lambda(x)$}{Lambda(x)} system: Wormhole spacetimes without nonlinear electrodynamics in unimodular gravity}

\author{G. Alencar}
\email{geova@fisica.ufc.br} 
\affiliation{Departamento de F\'isica, Universidade Federal do Cear\'a, Caixa Postal 6030, Campus do Pici, 60455-760 Fortaleza, Cear\'a, Brazil}

\author{T. M. Crispim}
\email{tiago.crispim@fisica.ufc.br}
\affiliation{Departamento de F\'isica, Universidade Federal do Cear\'a, Caixa Postal 6030, Campus do Pici, 60455-760 Fortaleza, Cear\'a, Brazil}

\begin{abstract}
In General Relativity, supporting known traversable wormhole geometries with electromagnetic fields typically requires complex Non-Linear Electrodynamics (NED). We demonstrate that Unimodular Gravity (UG) elegantly resolves this limitation. By relaxing energy-momentum conservation, UG introduces a dynamical cosmological term, $\Lambda(x)$, enabling a semi-classical energy exchange between matter and the vacuum. Exploiting this mechanism, we show how established exact wormhole spacetimes can be fully supported by a phantom scalar field coupled to standard linear Maxwell electrodynamics. We demonstrate that, provided the shape function $b(r)$ satisfies specific geometric conditions, this non-conservative framework naturally accommodates these solutions. By providing a new physical matter source mechanism that entirely bypasses the need for NED, our main contribution highlights UG as a powerful framework for sustaining established non-trivial topologies with simplified, well-understood classical fields.
\end{abstract}

\maketitle

\section{Introduction}

Unimodular Gravity (UG) stands as a robust alternative to General Relativity (GR) \cite{Smolin:2009ti,Bonder:2018mfz}; crucially, recent developments have established its inherent compatibility with the Einstein Equivalence Principle through the restriction of a fixed volume element $g=g_0$ \cite{3122468}. In this framework, first evoked by Einstein in 1919 \cite{cite_einstein_1919_if_you_have_it}, the field equations take the traceless form \cite{Smolin:2009ti},
\begin{equation}
R_{\mu\nu} - \frac{R}{4} g_{\mu\nu} = \kappa^2 \left(T_{\mu\nu} - \frac{1}{4}g_{\mu\nu}T\right), \label{EQUG}
\end{equation}
where the stress-energy tensor is defined in the usual way as 
\begin{equation}
    T_{\mu\nu} \equiv -2\frac{\delta L_m}{\delta g^{\mu\nu}}+g_{\mu\nu} L_m,
\end{equation}
with trace $T\equiv T^{\mu}_{\mu}$. Imposing the standard conservation law $\nabla_\mu T^{\mu\nu}=0$, the Bianchi identities applied to Eq.~\eqref{EQUG} trivially yield 
 \begin{equation}
  \partial_\mu(R+\kappa^2 T)=0   .
 \end{equation}
This leads to $R+\kappa^2 T=4\Lambda$, making UG classically equivalent to GR with a cosmological constant.

However, restricting metric variations to preserve the volume form ($g^{\mu\nu}\delta g_{\mu\nu}=0$) breaks full diffeomorphism symmetry down to volume-preserving ones, generated by divergence-free vectors ($\nabla_\mu \xi^\mu = 0$). This fundamentally alters the conservation paradigm \cite{Josset:2016vrq,Corral:2018hxi}. Varying a matter action $S_m$ along these restricted fields yields
\begin{equation}
\delta S_m = -\int T_{\mu\nu}\nabla^{\mu}\xi^{\nu}\sqrt{-g}\,d^4x = \int J_{\mu}\xi^{\mu}\sqrt{-g}\,d^4x,
\end{equation}
where $J^\mu \equiv \nabla_\nu T^{\mu\nu}$ parametrizes non-conservation. For invariance ($\delta S_m=0$), $J_\mu$ must be orthogonal to all divergence-free vectors. In simply-connected spacetimes, this integrability ($dJ=0$) dictates that $J_\mu$ is an exact gradient of a scalar $Q$, yielding 
\begin{equation}
    \nabla_\nu T^{\mu\nu} = \nabla^\mu Q \neq 0.
\end{equation}
Feeding this into the trace equations and defining $\kappa^2 Q \equiv \Lambda(x)$ yields the generalized field equations:
\begin{subequations}\label{UG_equations}
\begin{align}
G_{\mu\nu} +g_{\mu\nu}\Lambda(x)&=\kappa^2T_{\mu\nu},
\\
\nabla_\mu T^{\mu\nu}&=\nabla^\nu \Lambda .
\end{align}
\end{subequations}

Semiclassically, vacuum fluctuations do not gravitate directly in this formalism \cite{Josset:2016vrq,Perez:2017krv}. Since $\Lambda(x)$ emerges as an integration function rather than a fundamental Lagrangian coupling, it is protected from massive quantum corrections \cite{Salvio:2024vfl}. This dynamical freedom has been actively explored in cosmology \cite{Josset:2016vrq,Corral:2020lxt} and astrophysics, including gravitational waves \cite{Fabris:2021kxt} and compact objects \cite{Astorga-Moreno:2019uin,Fabris:2023rfg,Bengochea:2024mmu}. 

Recently, UG successfully resolved geometric frustrations in compact objects, demonstrating that exact Regular Black Holes and Black Strings can be supported by standard linear Maxwell electrodynamics, completely bypassing Non-Linear Electrodynamics (NED) \cite{3137491,Alencar:2026oxy}. Motivated by this, we begin by examining UG in its non-conservative formulation coupled solely to a scalar field. This initial setup reveals a critical limitation: a strict mathematical degeneracy between the scalar potential $V(\phi)$ and the running cosmological term $\Lambda(x)$. Since these terms appear only as an effective sum ($V + \Lambda$) within the gravitational equations, they cannot be uniquely disentangled, preventing a full exploitation of UG's generalized conservation laws. 

To break this $V + \Lambda$ degeneracy and restore the physical individuality of the fields, the introduction of an additional matter sector becomes mandatory; a role elegantly filled by the electromagnetic field. Motivated by this necessity, we apply this scalar--Maxwell--$\Lambda(x)$ framework to the study of traversable wormholes \cite{Morris:1988cz}. Historically, supporting known wormhole geometries that extend beyond the standard Ellis-Bronnikov solution has required the use of ``exotic'' or highly complex sources \cite{Crispim:2024dgd,Crispim:2024lzf,Silva:2025dgd,deSSilva:2024gdc,Bronnikov:2021uta,Bronnikov:2017sgg,Canate:2024lks,Channuie:2025,Sharif:2012,Mazharimousavi:2016,Cordeiro:2024non,Junior:2025dyo,Rodrigues:2023vab}, most notably NED, to satisfy the demanding structural requirements of these spacetimes. In contrast, our approach demonstrates that the non-conservative UG framework can naturally sustain these established geometries by entirely replacing NED with standard linear Maxwell electrodynamics.

In this work, we demonstrate that the modified conservation laws of UG establish a novel field interplay where the spatial gradient $\Lambda'(r)$ acts as an effective local source for the electromagnetic sector. This semi-classical energy exchange mechanism allows us to support exact wormhole geometries, specifically a class of shape functions characterized by power-law behaviors, a family of models extensively systematized by Lobo \cite{Lobo_cite}. By obtaining Lobo's  model without the need for NED or ad-hoc phenomenological sources, we show that UG provides a more fundamental and less contrived path for supporting complex traversable geometries.

The paper is organized as follows. Section \ref{sec:motivation} mathematically demonstrates how the non-conservative framework breaks the $V+\Lambda$ degeneracy via sector-wise conservation laws. Section \ref{sec:phi0_ned_noncons} establishes the exact formulation for $\Phi=0$ wormholes, deriving the modified field equations. In Sec.~\ref{aplicacao}, we apply this machinery to exact solutions: testing the Generalized Ellis--Bronnikov (GEB) geometry as a counter-example, and successfully constructing exact Maxwell wormholes for a constant shape function, $b(r)=b_0$, and power-law geometries. In Sec.\ref{secenergy}, we analyze the consequences of this framework for the energy conditions. We conclude in Sec.~\ref{sec:conclusions}. We adopt the metric signature $(-,+,+,+)$.

\section{Motivation: Breaking the Degeneracy Between the Scalar Potential and the Cosmological Sector}
\label{sec:motivation}

Before constructing wormhole solutions, it is important to clarify a structural issue that arises when a scalar field is coupled to a spacetime-dependent cosmological term. In particular, if the matter content consists only of a scalar field, the field equations depend only on an effective combination of the scalar potential and the cosmological sector. As a consequence, the scalar potential and the vacuum contribution cannot be separated unambiguously. This motivates the introduction of an additional matter component, which we take to be the electromagnetic field.

\subsection{Scalar field with a running cosmological term}
The effective field equations are given by
\begin{subequations}\label{UG_equations}
\begin{align}
\label{einsteinequg}G_{\mu\nu} +g_{\mu\nu}\Lambda(x)&=\kappa^2T_{\mu\nu},
\\
\nabla_\mu T^{\mu\nu}&=\nabla^\nu \Lambda .
\end{align}
\end{subequations}

Let the matter sector initially consist of a scalar field with Lagrangian
\begin{equation}
\mathcal{L}_\phi
=
-\epsilon (\nabla\phi)^2 - V(\phi),
\label{eq:scalar_lagrangian}
\end{equation}
where $\epsilon=+1$ corresponds to a canonical scalar field and $\epsilon=-1$ to a phantom scalar field.

The corresponding energy--momentum tensor is
\begin{equation}
T_{\mu\nu}^{(\phi)}
=
2\epsilon \nabla_\mu\phi \nabla_\nu\phi
-
g_{\mu\nu}\left[\epsilon(\nabla\phi)^2 + V(\phi)\right].
\label{eq:Tscalar}
\end{equation}

Its covariant divergence satisfies
\begin{equation}
\nabla^\mu T_{\mu\nu}^{(\phi)}
=
\left(2\epsilon \Box\phi - V_{,\phi}\right)\nabla_\nu\phi.
\label{eq:divTscalar}
\end{equation}

If the non-conservation of the matter sector is sourced by the spacetime dependence of $\Lambda(x)$, then the Bianchi identities imply
\begin{equation}
\nabla^\mu T_{\mu\nu}
=
\frac{1}{\kappa^2}\nabla_\nu\Lambda .
\label{eq:noncons}
\end{equation}

When only the scalar field is present, Eqs.~\eqref{eq:divTscalar} and \eqref{eq:noncons} give
\begin{equation}
2\epsilon \Box\phi
-
\left(
V + \frac{\Lambda}{\kappa^2}
\right)_{,\phi}
=0.
\label{eq:scalar_modified}
\end{equation}
Hence the scalar dynamics depends only on the effective combination
\begin{equation}
V_{\mathrm{eff}}(\phi) = V(\phi) + \frac{\Lambda(\phi)}{\kappa^2}.
\label{eq:Veff}
\end{equation}

The same degeneracy is present in the gravitational sector. Indeed, both
$V(\phi)$ and $\Lambda$ contribute proportionally to $g_{\mu\nu}$ and therefore appear only through the same effective combination in the Einstein equations.
The degeneracy can also be seen directly from the trace of Einstein equations. Taking the trace of Eq.~\eqref{einsteinequg} in four dimensions, one obtains
\begin{equation}
-R + 4\Lambda
=
\kappa^2 T.
\label{eq:trace_general}
\end{equation}

For the scalar field alone, the trace is
\begin{equation}
T^{(\phi)}
=
g^{\mu\nu}T_{\mu\nu}^{(\phi)}
=
-2\epsilon (\nabla\phi)^2 - 4V(\phi).
\label{eq:trace_scalar}
\end{equation}
Therefore Eq.~\eqref{eq:trace_general} becomes
\begin{equation}
R - 2\kappa^2 \epsilon (\nabla\phi)^2
=
4\left(\kappa^2 V + \Lambda\right).
\label{eq:trace_scalar_only}
\end{equation}

Once again, only the combination $\kappa^2 V + \Lambda$ appears. Therefore, with a scalar field alone, the transformations
\begin{equation}
V(\phi)\rightarrow V(\phi)+f(\phi),
\qquad
\Lambda(\phi)\rightarrow \Lambda(\phi)-\kappa^2 f(\phi),
\label{eq:degeneracy_shift}
\end{equation}
leave the system invariant. The scalar field by itself is thus unable to distinguish a genuine scalar potential from a vacuum contribution.

\subsection{Adding an electromagnetic sector}

To overcome this difficulty, we introduce an electromagnetic sector. For the sake of simplicity and to clearly illustrate the mechanism, we initially consider the linear Maxwell field, although the framework naturally extends to NED as we shall explore later. The Lagrangian is given by
\begin{equation}
\mathcal{L}_{M}
=
-\frac14 F_{\mu\nu}F^{\mu\nu},
\label{eq:Maxwell_lagrangian}
\end{equation}
whose energy--momentum tensor is
\begin{equation}
T_{\mu\nu}^{(M)}
=
F_{\mu\alpha}F_{\nu}{}^\alpha
-\frac14 g_{\mu\nu}F_{\alpha\beta}F^{\alpha\beta}.
\label{eq:Tmaxwell}
\end{equation}

A crucial property of Maxwell theory in four dimensions is that its energy--momentum tensor is traceless,
\begin{equation}
T^{(M)}=0.
\label{eq:trace_maxwell}
\end{equation}

We now consider the total matter content
\begin{equation}
T_{\mu\nu}
=
T_{\mu\nu}^{(\phi)}+T_{\mu\nu}^{(M)}.
\label{eq:Ttotal}
\end{equation}

The Einstein equations then read
\begin{equation}
G_{\mu\nu}+\Lambda(x) g_{\mu\nu}
=
\kappa^2 \left(T_{\mu\nu}^{(\phi)}+T_{\mu\nu}^{(M)}\right).
\label{eq:Einstein_total}
\end{equation}

\subsection{Trace equation for the scalar--Maxwell--\texorpdfstring{$\Lambda(x)$}{Lambda(x)} system}

Let us now examine the trace equation for the full system. Since the Maxwell sector is traceless, the total trace is simply
\begin{equation}
T
=
T^{(\phi)}+T^{(M)}
=
-2\epsilon (\nabla\phi)^2 - 4V(\phi).
\label{eq:trace_total}
\end{equation}

Therefore, taking the trace of Eq.~\eqref{eq:Einstein_total}, we obtain
\begin{equation}
-R+4\Lambda
=
\kappa^2\left[-2\epsilon (\nabla\phi)^2 - 4V(\phi)\right],
\end{equation}
or equivalently,
\begin{equation}
\Lambda
=
\frac{R}{4}
-
\frac{\kappa^2}{2}\epsilon(\nabla\phi)^2
-
\kappa^2 V(\phi).
\label{eq:Lambda_from_trace}
\end{equation}

This relation is extremely important. Although the Maxwell sector does not contribute directly to the trace, it modifies the geometry and therefore affects the Ricci scalar $R$. In this way, once the scalar field profile and the scalar potential are known, the trace equation determines $\Lambda$ separately.

Thus, unlike the purely scalar case, the system composed of a scalar field, an electromagnetic sector, and a running cosmological term is not governed solely by the combination $V+\Lambda/\kappa^2$. Instead, the scalar equation determines $V$, while the trace relation determines $\Lambda$.

\subsection{Sector-wise conservation laws}

We now assume that the scalar field remains individually conserved,
\begin{equation}
\nabla^\mu T_{\mu\nu}^{(\phi)}=0,
\label{eq:scalar_conserved}
\end{equation}
so that it obeys the standard equation of motion
\begin{equation}
2\epsilon \Box\phi - V_{,\phi}=0.
\label{eq:scalar_standard}
\end{equation}

As established in the previous subsections, the electromagnetic sector is required mathematically to break the structural degeneracy between the scalar potential $V(\phi)$ and the running cosmological term $\Lambda(x)$. Having introduced this independent sector, the deliberate phenomenological choice to route the non-conservation exclusively through the electromagnetic field---while keeping the scalar field separately conserved---is driven by two key physical motivations.

First, the scalar field responsible for supporting the wormhole throat is a phantom field ($\epsilon = -1$). Phantom fields are well known to be prone to severe classical and quantum instabilities due to their negative kinetic energy \cite{Carroll:2003st, Cline:2003gs}. If we allowed the scalar sector to be non-conserved, its equation of motion \eqref{eq:scalar_standard} would acquire an external driving or dissipative force term proportional to the gradient of the running vacuum, i.e., $\propto \nabla_\nu \Lambda$. For a phantom field, such an external coupling can easily trigger runaway kinetic instabilities, disrupting the stable, static configuration of the kink profile that holds the throat open.

Second, in semiclassical gravity and varying-$\Lambda$ cosmologies \cite{Freese:1986dd, Overduin:1998zv, Shapiro:2009dh}, a space-dependent cosmological term $\Lambda(x)$ represents a running vacuum energy density. The spatial variation or dynamical decay of this vacuum energy is physically expected to produce standard radiation, gauge bosons, or induce charge polarization, rather than generating exotic, negative-energy phantom particles \cite{Sola:2013gha}. Thus, it is physically more natural and thermodynamically consistent for the vacuum energy gradient $\nabla\Lambda$ to couple directly to the gauge field as a form of vacuum-to-radiation energy transfer, leaving the purely geometric scalar support intact.

The non-conservation required by Eq.~\eqref{eq:noncons} is then entirely carried by the electromagnetic sector,
\begin{equation}
\nabla^\mu T_{\mu\nu}^{(M)}
=
\frac{1}{\kappa^2}\nabla_\nu\Lambda.
\label{eq:EM_noncons}
\end{equation}

Using the standard identity
\begin{equation}
\nabla^\mu T_{\mu\nu}^{(M)}
=-
F_{\nu\alpha}\nabla_\mu F^{\mu\alpha},
\label{eq:Maxwell_identity}
\end{equation}
and defining an effective current through
\begin{equation}
\nabla_\mu F^{\mu\nu}=J^\nu,
\label{eq:Maxwell_with_current}
\end{equation}
we obtain
\begin{equation}
F_{\nu\alpha}J^\alpha
=-
\frac{1}{\kappa^2}\nabla_\nu\Lambda.
\label{eq:JLambda_relation}
\end{equation}

Equation~\eqref{eq:JLambda_relation} shows that a spacetime-dependent cosmological term acts as an effective source for the electromagnetic field.

\subsection{Relevance for wormhole geometries}

This framework is particularly convenient for static and spherically symmetric wormholes. In such geometries, the anisotropic stress required to support the throat can be provided by the scalar kinetic term, while the electromagnetic sector absorbs the non-conservation associated with the running cosmological term. The trace equation then allows one to reconstruct $\Lambda$ independently from the scalar potential.

In the following sections we apply this formalism to wormhole geometries and examine the interplay between the scalar, electromagnetic, and cosmological sectors in determining the spacetime structure.

\section{Scalar field, (non)linear electrodynamics, and non--conservation for \texorpdfstring{$\Phi=0$}{Phi=0} wormholes}
\label{sec:phi0_ned_noncons}

In this section we present the full construction of Morris--Thorne wormholes with vanishing redshift function, supported by a radial scalar field and a linear or NED sector in the presence of a non--conserved stress tensor.

\subsection{Metric, field content, and Einstein equations}

We consider the Morris--Thorne metric with vanishing redshift function \cite{Morris:1988cz},
\begin{equation}
ds^2
=
-dt^2
+
\frac{dr^2}{1-\frac{b(r)}{r}}
+
r^2\left(d\theta^2+\sin^2\theta\,d\varphi^2\right),
\label{eq:metric_phi0}
\end{equation}
where $b(r)$ is the shape function.

The Einstein equations are written as
\begin{equation}
G_{\mu\nu}+\Lambda(r)\,g_{\mu\nu}
=
\kappa^2
\left(
T^{(\phi)}_{\mu\nu}
+
T^{(\mathrm{NED})}_{\mu\nu}
\right),
\label{eq:einstein_phi0}
\end{equation}
where $\Lambda(r)$ is interpreted as an effective position--dependent cosmological sector.

The matter content consists of a radial scalar field $\phi=\phi(r)$, with Lagrangian \eqref{eq:scalar_lagrangian}, and a radial electric field
\begin{equation}
F_{tr}=E(r),\qquad F_{rt}=-E(r),
\end{equation}
described by a general Lagrangian
\begin{equation}
L=L(\mathcal F),\qquad
\mathcal F\equiv \frac14 F_{\mu\nu}F^{\mu\nu}.
\label{eq:ned_lag_phi0}
\end{equation}

\subsection{Scalar sector}

Since $\phi=\phi(r)$, and 
with the conventions used throughout this work, the scalar energy--momentum tensor reads
\begin{equation}
T^{(\phi)\mu}{}_{\nu}
=
-\epsilon\left(1-\frac{b(r)}{r}\right)\phi'(r)^2
\,\mathrm{diag}(1,-1,1,1)
-\delta^\mu{}_\nu\,V(\phi).
\label{eq:Tphi_phi0}
\end{equation}

Therefore,
\begin{align}
T^{(\phi)t}{}_{t}
&=
-\epsilon\left(1-\frac{b(r)}{r}\right)\phi'(r)^2-V(\phi),
\\[0.2cm]
T^{(\phi)r}{}_{r}
&=
+\epsilon\left(1-\frac{b(r)}{r}\right)\phi'(r)^2-V(\phi),
\\[0.2cm]
T^{(\phi)\theta}{}_{\theta}
=
T^{(\phi)\varphi}{}_{\varphi}
&=
-\epsilon\left(1-\frac{b(r)}{r}\right)\phi'(r)^2-V(\phi).
\end{align}

In particular,
\begin{equation}
T^{(\phi)r}{}_{r}-T^{(\phi)t}{}_{t}
=
2\epsilon\left(1-\frac{b(r)}{r}\right)\phi'(r)^2.
\label{eq:scalar_anisotropy}
\end{equation}

This is the anisotropic contribution that supports the wormhole throat.

Assuming that the scalar sector is conserved separately,
\begin{equation}
\nabla_\mu T^{(\phi)\mu}{}_{\nu}=0,
\label{eq:scalar_cons}
\end{equation}
its equation of motion is the standard one,
\begin{equation}
2\epsilon \Box\phi - V_{,\phi}=0.
\label{eq:scalar_eom_general}
\end{equation}

For the metric \eqref{eq:metric_phi0}, the scalar equation becomes
\begin{equation}
2\epsilon
\left[
\left(1-\frac{b(r)}{r}\right)\phi''
+
\frac{4r-3b(r)-r b'(r)}{2r^2}\,\phi'
\right]
-
V_{,\phi}=0.
\label{eq:scalar_eom_phi0}
\end{equation}

Multiplying by $\phi'(r)$, it is often convenient to rewrite it as
\begin{equation}
\begin{split}
V'(r) &= 2\epsilon\,\phi'(r) \Bigg[ \left(1-\frac{b(r)}{r}\right)\phi''(r) \\
&\quad + \frac{4r-3b(r)-r b'(r)}{2r^2}\,\phi'(r) \Bigg].
\end{split}
\label{eq:Vprime_general}
\end{equation}

This equation determines $V(r)$ once $\phi(r)$ is known.

\subsection{Electromagnetic sector}

For the general Lagrangian \eqref{eq:ned_lag_phi0}, the energy--momentum tensor is
\begin{equation}
T^{(\mathrm{NED})}_{\mu\nu}
=
g_{\mu\nu}L
-
L_{\mathcal F}\,F_{\mu\alpha}F_{\nu}{}^\alpha,
\qquad
L_{\mathcal F}\equiv \frac{dL}{d\mathcal F}.
\label{eq:Tned_cov}
\end{equation}

For a radial electric field,
\begin{equation}
\mathcal F
=
\frac14 F_{\mu\nu}F^{\mu\nu}
=
-\frac12\left(1-\frac{b(r)}{r}\right)E(r)^2.
\label{eq:F_phi0}
\end{equation}

Therefore,
\begin{align}
T^{(\mathrm{NED})t}{}_{t}
=
T^{(\mathrm{NED})r}{}_{r}
&=
L-2\mathcal F L_{\mathcal F},
\\[0.2cm]
T^{(\mathrm{NED})\theta}{}_{\theta}
=
T^{(\mathrm{NED})\varphi}{}_{\varphi}
&=
L.
\end{align}

Finally, the trace is
\begin{equation}
T^{(\mathrm{NED})}=4L-4\mathcal F L_{\mathcal F}.
\label{eq:trace_ned}
\end{equation}

\subsection{Einstein equations and master relations}

For the metric \eqref{eq:metric_phi0}, the Einstein tensor components are
\begin{equation}
G^t_t=-\frac{b'(r)}{r^2},
\,
G^r_r=-\frac{b(r)}{r^3},
\,
G^\theta_\theta
=
G^\varphi_\varphi
=
\frac{b(r)-r b'(r)}{2r^3}.
\label{eq:Gphi0}
\end{equation}

The Einstein equations \eqref{eq:einstein_phi0} then become
\begin{eqnarray}
&-&\frac{b'(r)}{r^2}+\Lambda(r) = \kappa^2 \bigg[ -\epsilon\left(1-\frac{b(r)}{r}\right)\phi'(r)^2 \nonumber \\
&-& V(r) + L(r) - 2\mathcal F(r)L_{\mathcal F}(r) \bigg], \label{eq:Ein_tt} \\[1em]
&-&\frac{b(r)}{r^3}+\Lambda(r) = \kappa^2 \bigg[ +\epsilon\left(1-\frac{b(r)}{r}\right)\phi'(r)^2 \nonumber \\
&-& V(r) + L(r) - 2\mathcal F(r)L_{\mathcal F}(r) \bigg], \label{eq:Ein_rr} \\[1em]
&&\frac{b(r)-r b'(r)}{2r^3}+\Lambda(r) = \kappa^2 \bigg[ -\epsilon\left(1-\frac{b(r)}{r}\right)\phi'(r)^2 \nonumber \\
&-& V(r) + L(r) \bigg]. \label{eq:Ein_ttang}
\end{eqnarray}

\subsubsection{Scalar reconstruction from $G^r{}_r-G^t{}_t$}

Subtracting \eqref{eq:Ein_tt} from \eqref{eq:Ein_rr}, the NED and $\Lambda$ contributions cancel:
\begin{equation}
\frac{b'(r)}{r^2}-\frac{b(r)}{r^3}
=
2\kappa^2\epsilon\left(1-\frac{b(r)}{r}\right)\phi'(r)^2.
\end{equation}

Thus,
\begin{equation}
\phi'(r)^2
=
\frac{1}{2\kappa^2\epsilon\left(1-\frac{b(r)}{r}\right)}
\left(
\frac{b'(r)}{r^2}-\frac{b(r)}{r^3}
\right).
\label{eq:phi_master}
\end{equation}

This shows that the scalar profile is fixed directly by the geometry.

\subsubsection{Electromagnetic master equation from $G^\theta{}_\theta-G^t{}_t$}

Subtracting \eqref{eq:Ein_tt} from \eqref{eq:Ein_ttang}, the scalar and $\Lambda$ sectors cancel:
\begin{equation}
\frac{b(r)-r b'(r)}{2r^3}+\frac{b'(r)}{r^2}
=
2\kappa^2\,\mathcal F(r)L_{\mathcal F}(r).
\end{equation}

Hence,
\begin{equation}
\mathcal F(r)L_{\mathcal F}(r)
=
\frac{b(r)+r b'(r)}{4\kappa^2 r^3}.
\label{eq:ned_master}
\end{equation}

This is the central equation for the NED sector.

It is convenient to define
\begin{equation}
H(r)\equiv \frac{b(r)+r b'(r)}{4\kappa^2 r^3},
\label{eq:Hdef}
\end{equation}
so that Eq.~\eqref{eq:ned_master} becomes
\begin{equation}
\mathcal F(r)L_{\mathcal F}(r)=H(r).
\label{eq:ned_master_H}
\end{equation}

\subsubsection{Trace equation}

Taking the trace of \eqref{eq:einstein_phi0},
\begin{equation}
-R+4\Lambda(r)
=
\kappa^2
\left(
T^{(\phi)}+T^{(\mathrm{NED})}
\right),
\label{eq:trace_full}
\end{equation}
with
\begin{eqnarray}
    T^{(\phi)}
&=&
-4V(r)-2\epsilon\left(1-\frac{b(r)}{r}\right)\phi'(r)^2,\\
T^{(\mathrm{NED})}&=&4L(r)-4\mathcal F(r)L_{\mathcal F}(r).
\end{eqnarray}

For $\Phi=0$,
\begin{equation}
R(r)=\frac{2b'(r)}{r^2}.
\label{eq:Rphi0}
\end{equation}

Using \eqref{eq:phi_master} and \eqref{eq:ned_master_H}, Eq.~\eqref{eq:trace_full} simplifies drastically to
\begin{equation}
L(r)=\frac{\Lambda(r)}{\kappa^2}+V(r).
\label{eq:Ltrace_final}
\end{equation}

This relation will be crucial below.

\subsection{Consistency of all Einstein equations}

At this point, one may ask whether Eqs.~\eqref{eq:phi_master}, \eqref{eq:ned_master_H}, and \eqref{eq:Ltrace_final} are sufficient to guarantee that the full Einstein system is satisfied. The answer is yes.

Indeed, define
\begin{equation}
A(r)\equiv \epsilon\left(1-\frac{b(r)}{r}\right)\phi'(r)^2.
\end{equation}

From \eqref{eq:phi_master},
\begin{equation}
2\kappa^2 A(r)
=
\frac{b'(r)}{r^2}-\frac{b(r)}{r^3},
\label{eq:A_relation}
\end{equation}
while from \eqref{eq:ned_master_H},
\begin{equation}
2\kappa^2 H(r)
=
\frac{b(r)+r b'(r)}{2r^3}.
\label{eq:H_relation}
\end{equation}

Using \eqref{eq:Ltrace_final}, the three Einstein equations reduce identically to their geometric left--hand sides. Therefore the system is fully consistent once the scalar field, the NED master equation, and the trace relation are simultaneously satisfied.

\subsection{Non--conservation and the modified NED field equation}

The non--conservation is assumed to be entirely carried by the NED sector,
\begin{equation}
\nabla_\mu T^{(\mathrm{NED})\mu}{}_{\nu}
=
\frac{1}{\kappa^2}\nabla_\nu \Lambda(r).
\label{eq:noncons_ned}
\end{equation}

The NED field equation becomes
\begin{equation}
\nabla_\mu\left(L_{\mathcal F}F^{\mu\nu}\right)=J^\nu.
\label{eq:ned_field_eq}
\end{equation}

Using
\begin{equation}
\nabla_\mu T^{(\mathrm{NED})\mu}{}_{\nu}
=-
F_{\nu\alpha}J^\alpha,
\end{equation}
we obtain
\begin{equation}
F_{\nu\alpha}J^\alpha
=
-\frac{1}{\kappa^2}\nabla_\nu\Lambda(r).
\label{eq:JLambda}
\end{equation}

For the radial electric ansatz, only the temporal component is nonvanishing, and one finds
\begin{equation}
J^t(r)=\frac{\Lambda'(r)}{\kappa^2 E(r)},
\qquad
J^r=J^\theta=J^\varphi=0.
\label{eq:Jt}
\end{equation}

The dynamical equation \eqref{eq:ned_field_eq} then becomes
\begin{equation}
\frac{d}{dr}
\left[
r^2\sqrt{1-\frac{b(r)}{r}}\,
L_{\mathcal F}(r)\,E(r)
\right]
=
\frac{r^2}{\kappa^2\sqrt{1-\frac{b(r)}{r}}}\,
\frac{\Lambda'(r)}{E(r)}.
\label{eq:ned_dyn}
\end{equation}

This is the equation that distinguishes the non--conservative framework from the conservative one.

\subsection{Construction algorithm for Maxwell electrodynamics}

The complete construction of $\Phi=0$ wormholes supported by a scalar field, Maxwell electrodynamics, and a non--conservative framework proceeds through a simplified sequence of steps. First, it is crucial to establish the geometric constraints required for a physically viable solution. The choice of the shape function $b(r)$ is restricted by two main conditions:
\begin{itemize}
    \item $1 - \frac{b(r)}{r} > 0$ for $r > r_0$: This is the standard wormhole requirement to ensure the absence of event horizons and a well-behaved spatial metric.
    \item $b(r) + r b'(r) > 0$ for $r > r_0$: In the context of our field equations, this term frequently governs the regularity of the electromagnetic sector. Ensuring it is strictly positive avoids naked singularities and prevents the fields from acquiring imaginary values outside the throat.
\end{itemize}

With a suitable geometry in hand, the algorithm for the exact Maxwell limit is given by:

\begin{enumerate}
    \item Fix the geometry through a shape function $b(r)$ satisfying the aforementioned constraints.
    
    \item Determine the scalar field derivative $\phi'(r)$ from Eq.~\eqref{eq:phi_master}.
    
    \item Integrate the scalar equation \eqref{eq:scalar_eom_phi0} to obtain the potential $V(r)$.
    
    \item Compute the geometric function $H(r)$ from Eq.~\eqref{eq:Hdef}.
    
  \item Impose the standard Maxwell Lagrangian, $L(\mathcal{F}) = -\mathcal{F}$, which fixes $L_{\mathcal{F}} = -1$. Using the electromagnetic invariant for this background geometry, $\mathcal{F}(r) = -\frac{1}{2}\left(1 - \frac{b(r)}{r}\right)E^2(r)$, the master equation \eqref{eq:ned_master_H} simplifies significantly to
  \begin{equation}
      \frac{1}{2}\left(1 - \frac{b(r)}{r}\right)E^2(r) = H(r).
  \end{equation}
   This allows us to determine the electric field directly as a function of the geometry:
    \begin{equation}
        E(r) = \sqrt{\frac{2H(r)}{1 - \frac{b(r)}{r}}}.
    \end{equation}
    
    \item Obtain the effective cosmological term from Eq.~\eqref{eq:Ltrace_final}. Evaluating the Maxwell Lagrangian density $L(r) = -\mathcal{F}(r) = \frac{1}{2}\left(1 - \frac{b(r)}{r}\right)E^2(r)$, this expression reduces to:
    \begin{equation}
        \Lambda(r) = \kappa^2 \left[ \frac{1}{2}\left(1 - \frac{b(r)}{r}\right)E^2(r) - V(r) \right].
    \end{equation}
    
    \item Finally, determine the effective non-conservative current from Eq.~\eqref{eq:Jt}. Furthermore, for the system to be consistent, Eq.~\eqref{eq:ned_dyn} must be satisfied. By setting the limit $\mathcal{L}_{\mathcal{F}} = -1$, this consistency condition reduces to
    \begin{equation}
        -\frac{d}{dr} \left[ r^2\sqrt{1-\frac{b(r)}{r}}\,E(r) \right] = \frac{r^2}{\kappa^2\sqrt{1-\frac{b(r)}{r}}}\, \frac{\Lambda'(r)}{E(r)}. \label{eq:ned_dyn_maxwell}
    \end{equation}
   
\end{enumerate}

This adapted framework provides a systematic and direct route to construct exact non--conservative Maxwell wormholes. Notably, it bypasses the complexities of inverting non-linear electrodynamics relations, showing that even standard linear electrodynamics can support these geometries when an energy exchange mechanism is allowed.

\section{Scalar--Maxwell--\texorpdfstring{$\Lambda(x)$}{Lambda(x)} wormholes}\label{aplicacao}
\subsection{A counter-example: The Generalized Ellis--Bronnikov geometry}

To illustrate the importance of the geometric constraints outlined in our construction algorithm, particularly for the Maxwell limit, it is instructive to analyze a geometry that fails to satisfy them. We consider the generalized Ellis--Bronnikov (GEB) shape function \cite{Kar:1995jz,Crispim:2024dgd}:
\begin{equation}
    b(r) = r - r^{3-2m}\left(r^m-b_0^m\right)^{2-\frac{2}{m}}, \qquad m>2.
    \label{eq:b_geb}
\end{equation}

It is useful to define the auxiliary function
\begin{equation}
    s(r) \equiv 1-\frac{b_0^m}{r^m}, \qquad \text{such that} \qquad 1-\frac{b(r)}{r}=s(r)^{2-\frac{2}{m}}.
    \label{eq:s_def}
\end{equation}

Substituting \eqref{eq:b_geb} into the scalar field master equation \eqref{eq:phi_master}, one finds
\begin{equation}
    \phi'(r)^2 = -\frac{m-1}{\kappa^2\epsilon\,r^2}\, \frac{1-s(r)}{s(r)} = \frac{(m -1)}{\epsilon r^2(1 - r^m/b_0^m)}.
    \label{eq:A_geb}
\end{equation}
For a phantom field ($\epsilon=-1$), Eq.~\eqref{eq:A_geb} is manifestly positive, and the scalar field solution is perfectly well-behaved \cite{Crispim:2024dgd}:
\begin{equation}
    \phi(r) = \phi_0 + 2\frac{\sqrt{m-1}}{m}\arctan\left(\frac{\sqrt{r^m - b_0^m}}{b_0^{m/2}}\right).
\end{equation}

However, the geometric viability for linear electrodynamics breaks down when we evaluate the master function $H(r)$. Substituting the shape function into the definition of $H(r)$, we obtain explicitly:
\begin{equation}
\begin{split}
    H_{\mathrm{GEB}}(r) = \frac{1}{2\kappa^2 r^2} \bigg\{ 1 &- \left[ 1 - (2-m)\left(\frac{b_0}{r}\right)^m \right] \\
    &\times \left[ 1 - \left(\frac{b_0}{r}\right)^m \right]^{1-\frac{2}{m}} \bigg\}.
\end{split}
\label{eq:H_geb_explicit}
\end{equation}

To understand whether this geometry can be supported by Maxwell electrodynamics, we must follow Step 5 of our algorithm, which requires computing the electric field via 
\begin{equation}
    E(r) = \sqrt{\frac{2H_{\mathrm{GEB}}(r)}{s(r)^{2 - \frac{2}{m}}}}.
    \label{eq:E_field_geb}
\end{equation}

This strictly demands $H_{\mathrm{GEB}}(r) \ge 0$ globally.

Let us inspect the asymptotic behavior of $H_{\mathrm{GEB}}(r)$ as $r \to \infty$:
\begin{equation}
    H_{\mathrm{GEB}}(r) \sim \frac{(m-1)(2-m)}{2m\kappa^2}\, \frac{b_0^m}{r^{m+2}}, \qquad r\to\infty.
    \label{eq:H_inf}
\end{equation}

Because the GEB geometry is defined for $m > 2$, the factor $(2-m)$ is strictly negative. Consequently, $H_{\mathrm{GEB}}(r) < 0$ at spatial infinity. 

This geometric violation directly prevents the construction of a Maxwell wormhole with this shape function. Imposing the standard Lagrangian $L(\mathcal{F}) = -\mathcal{F}$ yields:
\begin{equation}
    E^2(r) = \frac{2H_{\mathrm{GEB}}(r)}{1 - \frac{b(r)}{r}} < 0 \quad \text{as} \quad r \to \infty.
\end{equation}

Thus, the required radial electric field becomes purely imaginary away from the throat. This counter-example clearly demonstrates that a regular scalar sector and a throat devoid of horizons are insufficient on their own; the restrictive condition on the derivatives of $b(r)$ strictly dictates which geometries can be reconstructed by linear electrodynamics, even when non--conservative energy exchanges are permitted.

\subsection{An exact Maxwell wormhole: The \texorpdfstring{$b(r) = b_0$}{b(r) = b0} toy model}

To conclusively demonstrate the viability of our non--conservative construction algorithm, we now apply it to the simplest conceivable wormhole geometry, defined by a constant shape function:
\begin{equation}
    b(r) = b_0.
    \label{eq:b_toy}
\end{equation}

Before evaluating the field equations, we must verify if this toy model satisfies all the fundamental Morris--Thorne conditions alongside our specific geometric constraints for linear electrodynamics. First, the throat is correctly located at $r=b_0$, where $b(b_0) = b_0$. The flare-out condition is trivially satisfied since $b'(b_0) = 0 < 1$. Furthermore, asymptotic flatness is preserved as $b(r)/r = b_0/r \to 0$ when $r \to \infty$. Most importantly for the electromagnetic sector, the strict positivity constraints are globally obeyed for all $r > b_0$:
\begin{equation}
    1 - \frac{b(r)}{r} = 1 - \frac{b_0}{r} > 0,
\end{equation}
and the crucial regularity condition yields a strictly positive constant:
\begin{equation}
    b(r) + r b'(r) = b_0 > 0.
\end{equation}

Because this fundamental condition does not flip sign at spatial infinity (unlike the GEB geometry), this model is a prime candidate for a Maxwell realization. Following our construction algorithm, we substitute Eq.~\eqref{eq:b_toy} into the scalar master equation \eqref{eq:phi_master} to find the scalar field profile:
\begin{equation}
    \phi'(r)^2 = \frac{b_0}{2\kappa^2 r^2 (r - b_0)}.
\end{equation}
Assuming a phantom signature ($\epsilon = -1$) to ensure the right-hand side is positive, we take the square root and integrate to obtain the exact analytical expression for the scalar field:
\begin{equation}
    \phi(r) = \phi_0 \pm \frac{\sqrt{2}}{\kappa} \arctan\left(\sqrt{\frac{r}{b_0} - 1}\right),
\end{equation}
where $\phi_0$ is an integration constant. This yields a perfectly well-behaved, non-singular real scalar field $\phi(r)$ that holds the throat open, behaving as a typical kink profile. Integrating Eq.~\eqref{eq:scalar_eom_phi0} then furnishes a strictly positive scalar potential:
\begin{equation}
V(r) = \frac{b_0}{6\kappa^2 r^3} \implies V(\phi) = \frac{1}{6\kappa^2 b_0^2} \cos^6\left(\frac{\kappa \phi}{\sqrt{2}}\right).
\end{equation}

Next, we evaluate the geometric master function $H(r)$ via Eq.~\eqref{eq:Hdef}. For $b(r)=b_0$, it simplifies dramatically to:
\begin{equation}
    H_{\text{toy}}(r) = \frac{b_0}{4\kappa^2 r^3}.
\end{equation}
Since $H_{\text{toy}}(r) > 0$ globally, we can seamlessly impose the Maxwell limit ($L_{\mathcal{F}} = -1$) without encountering imaginary fields. From Step 5 of our algorithm, the radial electric field is exactly determined:
\begin{equation}
    E(r) = \sqrt{\frac{2 H_{\text{toy}}(r)}{1 - \frac{b_0}{r}}} = \frac{\sqrt{b_0}}{\sqrt{2}\kappa r \sqrt{r - b_0}}.
    \label{eq:E_field_toy}
\end{equation}
Remarkably, for this specific geometry, the magnitude of the electric field is perfectly proportional to the scalar field derivative. This field is real and finite everywhere outside the throat, decaying properly as $r \to \infty$.

Finally, the closure of the system requires the dynamic exchange of energy. The effective cosmological term acts as an energy reservoir, adapting to the geometry. It becomes:
\begin{equation}
    \Lambda(r) = \kappa^2 \left[ L(r) - V(r) \right] = \frac{b_0}{12 r^3}.
\end{equation}
The non--conservative current $J^t(r)$, which drives the necessary energy exchange to sustain the linear electromagnetic field, is then given by
\begin{equation}
    J^t(r) = -\frac{\sqrt{2b_0(r-b_0)}}{4\kappa r^3}.
\end{equation}

Notably, this result ensures that the consistency condition \eqref{eq:ned_dyn} is identically satisfied for the Maxwell limit $\mathcal{L}_{\mathcal{F}} = -1$, confirming the internal coherence of the unimodular reconstruction for this geometry.

This simple toy model holds profound physical implications. In the standard conservative framework, exact wormhole solutions supported by ordinary Maxwell electrodynamics are notoriously difficult to construct---often requiring exotic charge distributions, fine-tuning, or suffering from naked singularities. 

Here, by allowing the generalized energy-momentum tensor to be non--conservative ($\nabla_\mu T^{\mu\nu} \neq 0$), the electric field does not need to satisfy the rigid source-free Maxwell equations. Instead, the spatial variation of the effective cosmological term $\Lambda(r)$ continuously injects or absorbs energy from the electromagnetic sector via the effective current $J^t(r)$. It is precisely this energy channel that bypasses the restrictive no-go theorems of conservative Einstein-Maxwell-scalar systems, proving that standard linear electrodynamics is perfectly capable of supporting smooth wormhole geometries when embedded in a non--conservative background.

\subsection{Maxwell wormholes from power-law shape functions}

To further demonstrate the robustness of our non--conservative algorithm, we extend our analysis to a more general family of geometries defined by the power-law shape function \cite{Lobo_cite,Lobo:2005yv}
\begin{equation}
   \label{perfil} b(r) = b_0 (b_0/r)^\gamma,
\end{equation}
 with $0<\gamma < 1$. This family ensures asymptotic flatness while allowing for different rates of spatial curvature decay. 

For this generalized case, we again verify that the Morris--Thorne criteria and our geometric consistency requirements are satisfied. The throat remains situated at $r=b_0$, where $b(b_0) = b_0$. The flare-out condition is strictly obeyed since $b'(b_0) = -\gamma < 1$, ensuring the throat opens correctly for any $0 < \gamma < 1$. Furthermore, asymptotic flatness is preserved as $b(r)/r = (b_0/r)^{\gamma+1} \to 0$ when $r \to \infty$. Regarding the stability and regularity of the electromagnetic sector, the positivity constraints are globally maintained for all $r > b_0$:
\begin{equation}
1 - \frac{b(r)}{r} = 1 - \left(\frac{b_0}{r}\right)^{\gamma+1} > 0,
\end{equation}
and the crucial regularity term likewise yields a strictly positive profile:
\begin{equation}
b(r) + r b'(r) = (1-\gamma) b_0 \left(\frac{b_0}{r}\right)^\gamma > 0.
\end{equation}

Following the construction steps, we first determine the scalar field derivative from Eq.~\eqref{eq:phi_master} by substituting the explicit form of $b(r)$:
\begin{equation}
    \phi'(r)^2 = \frac{(1+\gamma) b_0^{\gamma+1}}{2\kappa^2 r^{\gamma+2} \left[ r - b_0 \left( \frac{b_0}{r} \right)^\gamma \right]}.
\end{equation}
Assuming a phantom signature ($\epsilon = -1$), the scalar field $\phi(r)$ is obtained by integration. For the general case, the solution can be expressed as:
\begin{multline}
    \phi(r) = \phi_0 + \frac{\sqrt{2}}{\kappa\sqrt{\gamma+1}}\arctan\left(\frac{\sqrt{r^{1+\gamma}-b_0^{1+\gamma}}}{b_0^{(1+\gamma)/2}}\right).
\end{multline}
Integrating Eq.~\eqref{eq:scalar_eom_phi0} for this geometry leads to the explicit scalar potential:

\begin{eqnarray}
  V(r) &=& \frac{(1-\gamma)(1+\gamma)}{2(3+\gamma)\kappa^2} \frac{b_0^{\gamma+1}}{r^{3+\gamma}}, \,\, \text{or} \\
    V(\phi) &=& \frac{(1-\gamma)(1+\gamma)}{2(3+\gamma)\kappa^2 b_0^2} \left[ \cos\left( \frac{\kappa \sqrt{\gamma + 1}}{\sqrt{2}} \phi \right) \right]^{\frac{2(3+\gamma)}{1+\gamma}}.
\end{eqnarray}

\begin{figure}[ht!]
    \centering
    \includegraphics[width=1\linewidth]{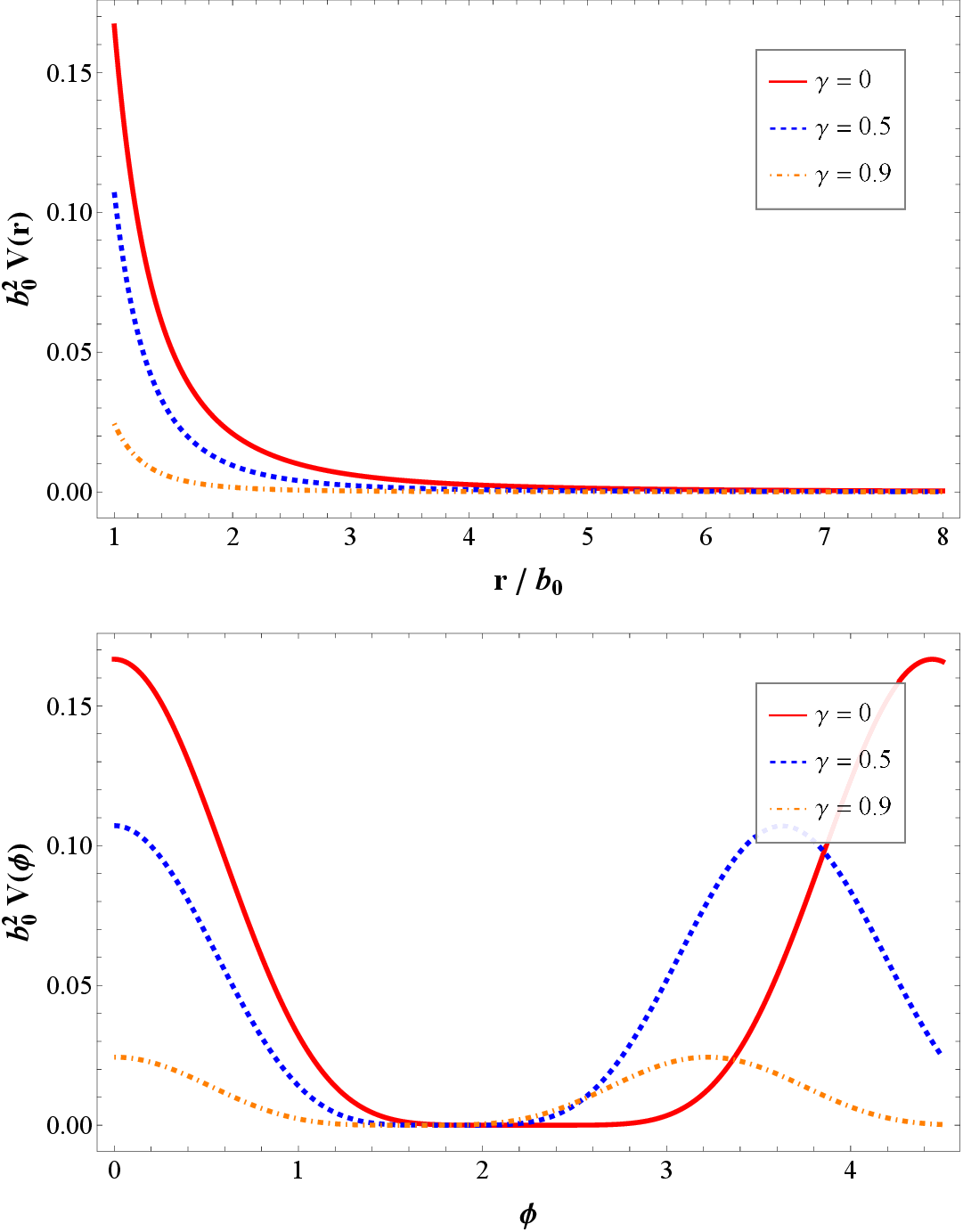}
    \caption{Plot of the scalar potential $V$ as a function of $r$ (top) and as a function of the scalar field $\phi$ (bottom), for different values of $\gamma$ and with $\kappa = 1$ and $\phi_0 = 0$ fixed.}
    \label{fig:V}
\end{figure}

In Fig.\ref{fig:V} we have the plot of the scalar potential $V$ as a function of $r$ (top) and as a function of the scalar field $\phi$ (bottom), for values of $\gamma = \{ 0, 0.5, 0.9\}$. From the plots, it can be seen that the potential decays asymptotically to zero as $r \to \infty$, while exhibiting periodic behavior as a function of the scalar field.

Next, we evaluate the geometric master function $H(r)$, which governs the electromagnetic sector:
\begin{equation}
    H(r) = \frac{(1-\gamma) b_0^{\gamma+1}}{4\kappa^2 r^{3+\gamma}}.
\end{equation}
Since $H(r) > 0$ for all $\gamma < 1$, the Maxwell limit ($L_{\mathcal{F}} = -1$) is globally accessible. The radial electric field $E(r)$ is then recovered explicitly as a function of $r$:
\begin{equation}
    E(r) = \sqrt{ \frac{(1-\gamma) b_0^{\gamma+1}}{2 \kappa^2 r^{2+\gamma} \left[ r - b_0 \left( \frac{b_0}{r} \right)^\gamma \right]} }.
\end{equation}

\begin{figure}[h]
    \centering
    \includegraphics[width=1\linewidth]{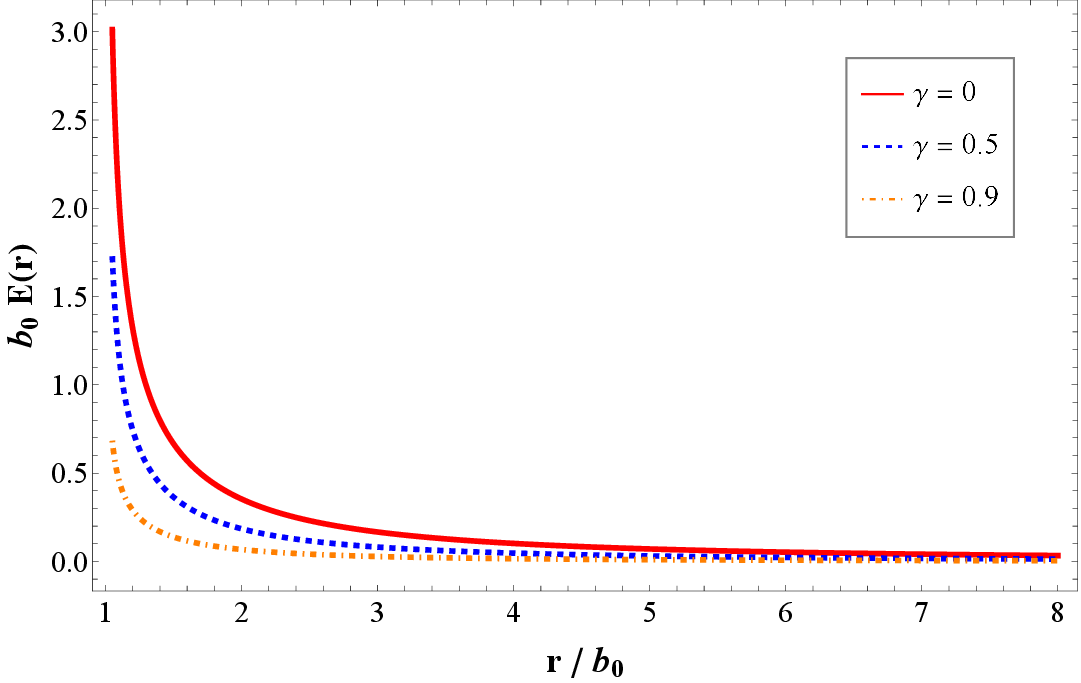}
    \caption{Plot of the electric field as a function of $r$ for different values of $\gamma$, with $\kappa$ fixed at 1.}
    \label{fig:E}
\end{figure}

In Fig.\eqref{fig:E}, we show the behavior of the electric field as a function of $r$, where one can observe its decay to zero in the limit $r \to \infty$, where spacetime is asymptotically flat. Due to the current $J^\mu$, it can be seen that the electric field does not exhibit the Coulombic behavior characteristic of conservative systems.

\begin{figure}
    \centering
    \includegraphics[width=1\linewidth]{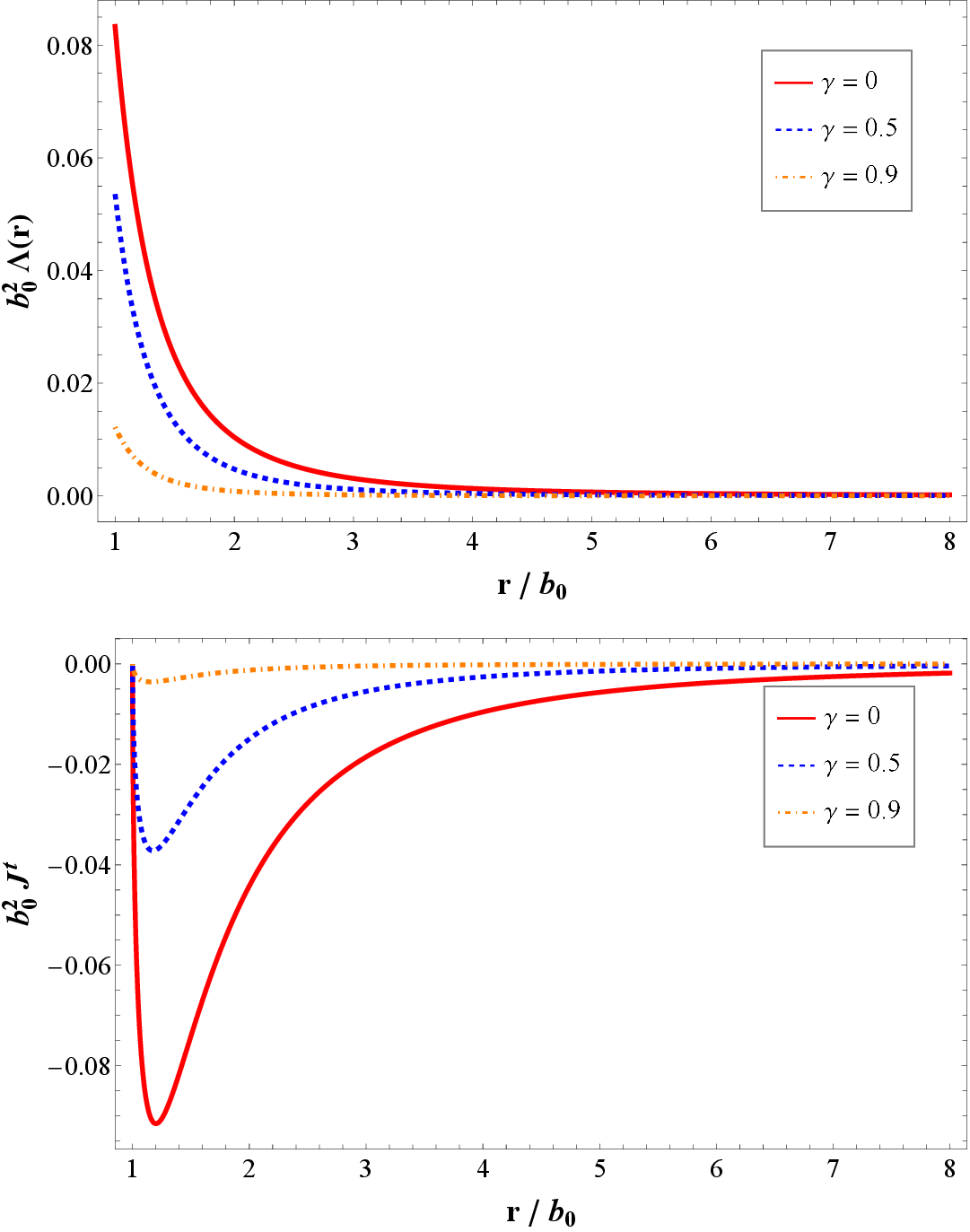}
    \caption{Plot of $\Lambda(r)$ (top) and the current $J^t(r)$ (bottom) as a function of $r$, for different values of $\gamma$.}
    \label{fig:JL}
\end{figure}

The effective cosmological term $\Lambda(r)$ acts as the energy reservoir for this system. By evaluating the difference between the Maxwell Lagrangian and the scalar potential, we find:
\begin{equation}
\label{vacuogamma} \Lambda(r) = \kappa^2[L(r) - V(r)] = \frac{(1-\gamma)^2}{4(3+\gamma)} \frac{b_0^{\gamma+1}}{r^{3+\gamma}}
\end{equation}
Finally, the non--conservative current $J^t(r)$ that sustains this Maxwellian configuration is determined by the spatial gradient of $\Lambda(r)$ and the electric field:
\begin{equation}
    J^t(r) = -\frac{(3+\gamma) \Lambda(r)}{\kappa^2 r} \sqrt{ \frac{2 \kappa^2 r^{2+\gamma} \left[ r - b_0 \left( \frac{b_0}{r} \right)^\gamma \right]}{(1-\gamma) b_0^{\gamma+1}} }.
\end{equation}
In Fig.\eqref{fig:JL} we have the plot of $\Lambda(r)$ (top) and the current $J^t(r)$ (bottom) as a function of $r$, for different values of $\gamma$. The physical behavior of the effective current $J^t$ and the running vacuum energy $\Lambda(r)$ reveals a consistent picture of the dynamic energy-exchange mechanism sustaining the wormhole geometry. As illustrated by their radial profiles, the vacuum energy density $\Lambda(r)$ monotonically decays from a maximum at the throat to zero at spatial infinity, ensuring that the spacetime remains asymptotically flat. Consequently, the negative radial gradient of the vacuum ($\Lambda'(r) < 0$) dynamically induces a strictly negative effective charge density $J^t$ to counterbalance the vacuum decay. Remarkably, this induced current does not peak exactly at the throat; rather, it reaches a minimum (maximal absolute magnitude) just outside it. This valley marks the region of maximum vacuum polarization and energy transfer, precisely where the gravitational gradients are most extreme. Exactly at the throat ($r = b_0$), $J^t$ vanishes completely, guaranteeing a regular geometric transition surface devoid of singular charge concentrations. At spatial infinity ($r \to \infty$), where the spacetime becomes flat and the vacuum stabilizes ($\Lambda' \to 0$), the current smoothly vanishes, naturally turning off the thermodynamic exchange. Furthermore, the parameter $\gamma$ acts as a continuous dial for this non-conservative mechanism: as $\gamma \to 1$, both $\Lambda(r)$ and $J^t$ are entirely suppressed, smoothly recovering the classical, conservative, and charge-free Ellis--Bronnikov limit.

Again, this result ensures that the consistency condition \eqref{eq:ned_dyn} is identically satisfied for the Maxwell limit $\mathcal{L}_{\mathcal{F}} = -1$, confirming the internal coherence of the unimodular reconstruction for this general geometry.

The existence of this continuous family of solutions proves that our non--conservative framework can support a wide range of power-law geometries, parameterized by $0 \leq \gamma < 1$. Notably, all physical quantities decay as power laws at spatial infinity, ensuring a well-behaved asymptotic limit. The lower bound $\gamma = 0$ perfectly recovers the constant shape function toy model analyzed previously. Even more remarkably, the exact limit $\gamma \to 1$ acts as a critical decoupling threshold, where the shape function becomes $b(r) \to b_0^2/r$, corresponding precisely to the classic Ellis--Bronnikov wormhole.

In this limit, the $(1-\gamma)$ factors deeply embedded in the geometric constraints elegantly deactivate the non-scalar sectors. The radial electric field and the effective cosmological term are driven to zero. Consequently, by virtue of the field equations' trace relation, the scalar potential is also identically suppressed, recovering a strictly massless field. Furthermore, a scaling analysis of the effective source current reveals that it vanishes proportionally to $(1-\gamma)^{3/2}$. This specific scaling guarantees a highly smooth thermodynamic shutoff of the energy exchange mechanism as the critical limit is approached. The system seamlessly reduces to a pure, massless phantom scalar field in vacuum, with the direct integration of the scalar equation naturally recovering the canonical arctangent profile of the Ellis--Bronnikov spacetime.

Physically, this demonstrates that the classic Ellis--Bronnikov geometry represents a singular, uncharged, conservative ``ground state'' where the necessary anisotropic stress to hold the throat open is fully self-sufficiently provided by the scalar kinetic term. For any intermediate value $0 \leq \gamma < 1$, the geometry is not self-sustaining in a conservative vacuum; it strictly requires the dynamic energy pump provided by $J^t(r)$ to continuously exchange energy between the unimodular vacuum and the Maxwell sector to stabilize the tangential pressures. Therefore, this non-conservative mechanism is precisely what enables standard linear electrodynamics to continuously deform the geometry away from the critical limit, bridging these well-known spacetimes and populating a regime that would be strictly forbidden in conservative GR.

\section{Energy conditions}\label{secenergy}
To fully assess the physical viability of the exact geometry \eqref{perfil} and the specific role played by each matter field, we perform an explicit evaluation of the standard energy conditions. In our framework, the geometric unimodular term acts as an effective vacuum energy distribution (Eq.\eqref{vacuogamma}),
\begin{equation}
    \Lambda(r) = \frac{(1-\gamma)^2}{4(3+\gamma)} \frac{b_0^{\gamma+1}}{r^{3+\gamma}} \equiv C(\gamma)\frac{b(r)}{r^3},
\end{equation}
which shifts the physical matter density and pressures. Using the Einstein equations, the effective matter quantities $\rho$, $p_r$, and $p_t$ for the combined scalar and exact Maxwell sector can be evaluated from
\begin{equation}
    T_\mu^\nu = \text{diag}(-\rho, p_r,p_t,p_t).
\end{equation}
To test the Null Energy Condition (NEC), we compute the fundamental combinations $\rho + p_i$. Remarkably, the effective vacuum term $\Lambda(r)$ cancels out identically in these combinations, leading to drastically simplified expressions:

\begin{eqnarray}\kappa^2 (\rho + p_r) &=& - (1 + \gamma) \frac{b(r)}{r^3}, \label{eq:nec_radial_gamma} \\\kappa^2 (\rho + p_t) &=& \frac{1 - \gamma}{2} \frac{b(r)}{r^3}. \label{eq:nec_tangential_gamma}\end{eqnarray}
These concise relations reveal a perfect decoupling of the physical roles of the fields. First, because $b(r) > 0$ and $\gamma > 0$, Eq.~\eqref{eq:nec_radial_gamma} is strictly negative. The radial NEC is fundamentally violated everywhere outside the throat, a standard requirement for traversable wormholes. Because the Maxwell limit ensures that $\rho_{\mathrm{Maxwell}} + p_{r, \mathrm{Maxwell}} = 0$, this violation is driven \textit{exclusively} by the kinetic term of the phantom scalar field. The violation is maximal at the throat ($r=b_0$), where the geometric factor $b(r)/r^3$ peaks, and decays asymptotically as $1/r^{3+\gamma}$. Furthermore, as $\gamma \to 1$, the asymptotic decay of the geometry is steeper, requiring a stronger concentration of exotic phantom matter near the throat to sustain the topology, which translates into a larger magnitude of $\rho + p_r \propto -(1+\gamma)$. Conversely, Eq.~\eqref{eq:nec_tangential_gamma} proves that the tangential NEC is globally satisfied, provided $\gamma < 1$. Because a purely radial scalar field satisfies $\rho_\phi + p_{t,\phi} = 0$, the tangential regularity is sustained \textit{exclusively} by the linear Maxwell electrodynamics. Since the radial NEC is violated, the Weak Energy Condition (WEC, $\rho \ge 0$) and Dominant Energy Condition (DEC, $\rho \ge \vert{}p_i\vert{}$) are inevitably violated as well. Indeed, the matter energy density, 
\begin{equation}
  \kappa^2 \rho = - [\gamma + C(\gamma)] b(r)/r^3  ,
\end{equation} is manifestly negative everywhere.

It is highly instructive to compare our exact field--theoretic construction with the standard phenomenological approaches to phantom energy wormholes, such as in Ref.~\cite{Lobo_cite}. In standard phantom wormholes, the exotic matter is modeled as a single perfect fluid governed by a postulated equation of state $p_r = \omega \rho$, restricted to the phantom regime ($\omega < -1$). In such models, the fluid is strictly conserved ($\nabla_\mu T^{\mu\nu} = 0$). To satisfy the field equations and shape function requirements, the fluid's energy density $\rho$ can often be chosen to be strictly positive ($\rho > 0$), with the NEC violation arising solely because the immense negative radial pressure overpowers the positive density ($\rho + p_r = \rho(1+\omega) < 0$).Our unimodular formulation operates under profoundly different physical principles:

\begin{enumerate}\item \textbf{No prescribed Equation of State:} We do not assume an EoS \textit{a priori}. The effective state relations emerge dynamically from the underlying field theory, specifically, from the interplay of a phantom scalar field and a standard Maxwell Lagrangian.\item \textbf{Negative Energy Density:} Unlike the phenomenological fluid where $\rho > 0$ is achievable, our model inherently possesses a globally negative matter density. This is an unavoidable physical consequence of the phantom scalar field ($\epsilon = -1$) dictating the dominant energetic contribution to support the geometry.\item \textbf{Dynamical Energy Exchange:} The most defining feature of our model is the breaking of local conservation. The specific $\gamma$--profile is maintained not by a tailored isolated fluid, but by an open thermodynamic channel. The effective current $J^\mu$ continuously mediates an energy exchange between the vacuum sector, encoded in the spatially varying $\Lambda(r)$, and the Maxwell electromagnetic sector. This energy flow allows standard linear electrodynamics to regularize the tangential pressures and support the wormhole without requiring ad hoc non--linear field distributions.\end{enumerate}

Finally, as previously stated, the limit $\gamma \to 1$ recovers the geometry and matter sources of the Ellis-Bronnikov wormhole. Therefore, in this limit, the combinations of densities and pressures smoothly reduce to those already known for this geometry, which has the scalar field's kinetic term as its sole matter source.

\section{Final remarks}
\label{sec:conclusions}

In this work, we have explored a novel and systematic approach to construct exact traversable wormhole geometries coupled to electromagnetic fields. In standard GR, the field equations for such configurations are notoriously rigid. The requirement of maintaining a regular throat without event horizons creates an overconstrained system when one attempts to introduce standard linear Maxwell electrodynamics. Historically, this geometric frustration forced the community to rely heavily on NED or ad-hoc anisotropic phenomenological fluids simply to close the system of equations.

We have demonstrated that UG, particularly when formulated with a non-conserved energy--momentum tensor, offers a fundamentally more elegant resolution. By restricting the spacetime symmetries to volume-preserving diffeomorphisms, the standard conservation law is relaxed, allowing the emergence of a spacetime-dependent cosmological term, $\Lambda(x)$. This mathematical structure provides a natural mechanism for energy exchange between the matter sector and the gravitational vacuum. Physically, this non-conservative framework aligns with the expectations of Quantum Field Theory in curved spacetimes, where the vacuum energy density is not a rigid constant but a dynamical parameter that can ``run'' with the local energy scale or gravitational curvature \cite{Shapiro:2009dh, Sola:2013gha}. In regions with steep gravitational gradients, such as the vicinity of a wormhole throat, the decay or variation of this vacuum energy dynamically induces a polarization response. This energy transfer manifests macroscopically as an effective current $J^\mu$ \cite{Freese:1986dd, Overduin:1998zv}, which acts not as a distribution of fundamental free charges, but as the exact physical mediator converting the local decay of the vacuum energy directly into electromagnetic energy to sustain the tangential pressures.

By exploiting this mechanism, we successfully constructed exact, analytical Scalar–Maxwell–$\Lambda(x)$ wormhole solutions. We applied this machinery to the GEB geometry and a broader family of power-law geometries. Crucially, the GEB case serves as a pedagogical benchmark to demonstrate that not all wormhole metrics are compatible with this framework; the shape function $b(r)$ must satisfy specific geometric conditions to allow for a physically consistent energy exchange. For the valid solutions, we proved that exact wormhole spacetimes can be perfectly supported by a phantom scalar field coupled to standard linear Maxwell electrodynamics. In this setup, isolating the non-conservation exclusively to the electromagnetic sector while keeping the scalar field strictly conserved is a deliberate and crucial phenomenological choice. Allowing the phantom scalar field ($\epsilon = -1$) to interact with the vacuum gradient would introduce external dissipative forces into its equation of motion, inevitably triggering runaway kinetic instabilities \cite{Carroll:2003st, Cline:2003gs}. By keeping the scalar sector minimally coupled and independently conserved, we protect the static geometric support of the throat. Thus, the standard Maxwell field safely absorbs the non-conservation, entirely bypassing the complexities of inverting non-linear Lagrangian relations, yielding a cleaner and more transparent matter sector that preserves the target metrics while highlighting the interplay between geometry and UG's generalized conservation laws.

From a broader perspective, these results suggest that exotic geometries, such as regular black holes and traversable wormholes, might not strictly require highly complex or unnatural matter sources if the underlying gravitational framework allows for semi-classical energy exchange with the vacuum. 

Future investigations must naturally extend this framework to address the dynamical viability of these solutions. A critical next step is to perform a rigorous perturbative stability analysis. It is well known in standard GR that traversable wormholes supported by phantom scalar fields generally suffer from instabilities under radial perturbations. However, we expect the stability properties in our non-conservative UG framework to be fundamentally altered. Because the geometry is coupled to an effective vacuum reservoir via $\Lambda(r)$, the non-conservative energy exchange could act as a dynamic feedback mechanism. Specifically, if the throat is subjected to a radial perturbation, the induced variations in the effective current $J^\mu$ might dampen the perturbation by absorbing kinetic energy from the fields or injecting restoring energy back into the electromagnetic sector, potentially quenching the standard phantom instabilities. However, at this stage, this phenomenon is purely speculative, since evaluating the exact spectrum of the perturbation equations to confirm whether this open thermodynamic channel provides a true stabilizing effect is a compelling subject for future work.

Ultimately, this work highlights UG not just as a tool for addressing the cosmological constant problem, but as a rich, effective framework for modeling compact objects and non-trivial spacetime topologies with simplified, well-understood classical fields.

\acknowledgments{We acknowledge the financial support provided by the Conselho Nacional de Desenvolvimento Científicoe Tecnológico (CNPq), Fundação Cearense de Apoio ao Desenvolvimento Científico e
Tecnológico (FUNCAP).   This
study was financed in part by the Coordenação de
Aperfeiçoamento de Pessoal de Nível Superior- Brasil
(CAPES)- Finance Code 001.
}
\bibliographystyle{apsrev4-2}
\bibliography{ref}
\end{document}